\documentclass[
	a4paper, 
	10pt, 
	unnumberedsections, 
	twoside, 
]{LTJournalArticle}
\usepackage{gensymb}
\usepackage{cite}
\usepackage{hyperref}
\usepackage{microtype}

\runninghead{Temperature Reconstruction} 

\footertext{\textit{preprint}} 

\title{Forensically useful mid-term and short-term temperature reconstruction for quasi-indoor death scenes} 

\author{%
    Jędrzej Wydra\textsuperscript{1,2}\thanks{Corresponding author: \href{mailto:jedrzej.wydra@amu.edu.pl}{jedrzej.wydra@amu.edu.pl}}, 
    Łukasz Smaga\textsuperscript{3} and 
    Szymon Matuszewski\textsuperscript{1,2}
}

\date{\footnotesize\textsuperscript{\textbf{1}}Laboratory of Criminalistics, Adam Mickiewicz University in Poznań, al. Niepodległości 53, Poznań 61-714, Poland\\ \textsuperscript{\textbf{2}}Centre for Advanced Technologies, Adam Mickiewicz University in Poznań, ul. Uniwersytetu Poznańskiego 10, Poznań 61-614, Poland\\ \textsuperscript{\textbf{3}}Department of Mathematical Statistics and Data Analysis, Adam Mickiewicz University in Poznań, ul. Uniwersytetu Poznańskiego 4, Poznań 61-614, Poland}

\renewcommand{\maketitlehookd}{%
	\begin{abstract}
		While estimating postmortem interval (PMI) ambient temperature plays pivotal role, so its reconstruction is crucial for forensic scientists. The recommended procedure is to correct temperatures from the nearest meteorological station based on measurements from the death scene; typically applying linear regression. Recently, there were attempts to use different algorithms, which can improve that correction, for example GAM algorithm. Unfortunately, the improvements are usually a consequence of using more dependent variables than just the temperature from the death scene (e.g. humidity), which is impractical.
  
        This study develops a practical new methods to accurately reconstruct ambient temperatures at a death scene, using just temperature measurements. Since the main difficulty preventing practitioners from using the correction protocol more frequently is likely the need to record temperatures on site for at least several days, we searched for the possibilities to shorten the measurement period. For this purpose we tested two less popular algorithms, which gave hope for this shortening. They were the concurrent regression model (the model from the functional data analysis field) for the mid-term reconstruction (measurements lasting several days) and the functional model based on Fourier expansion for the short-term reconstruction (measurements lasting a few hours).
        
        Performance of the algorithms was tested using data collected in six places: roof and attic of the heated building, unheated garage inside the heated building, unheated wooden shack, uninhabited building and underground (the data logger was buried about 30 cm below the ground level). We classified these places as quasi-indoor conditions in contrast to typical indoor conditions, where temperatures are nearly constant and typical outdoor conditions in case of which there is no heat insulation. 
        
        The mid-term model reduced error compared to the linear regression, providing nearly perfect reconstruction for measurement periods longer than six days. More importantly, however, the accuracy of short-term reconstruction was also high. The short-term model closely matched the concurrent regression model’s performance after only four to five hours of measurements.
        
        In practice, both methods are very similar to the standard procedure. The main difference is the change in algorithm and its implementation. In conclusion, this study demonstrates that correction of weather station temperatures can provide fairly accurate temperature data for use in estimating PMI after only 4-5 hours of measurements on a death scene.

	\end{abstract}

 \textbf{Keywords:} Forensic science, Post-mortem interval, Temperature reconstruction, Functional data analysis
}

\begin{document}

\maketitle 

\setlength{\parskip}{0pt}
\section{Introduction}

Temperature plays a pivotal role in estimating postmortem interval (PMI), given its significant impact on body decomposition dynamics and growth of insects that colonize cadavers \cite{higley2001insect, amendt2007best}. Consequently, establishing standards or guidelines for managing temperature data remain critical pursuit for forensic scientists, particularly within forensic entomology but also in other fields, for which temperature of a death scene is a key factor \cite{higley2001insect, amendt2007best, archer2004effect, dourel2010using, johnson2012experimental, martineau2017trabecular, michalski2018thermal, hofer2020estimating, jeong2020extended, lutz2020stay, moreau2021honey}. 

Some authors recommend correcting data from meteorological station based on temperature measurements at the death scene \cite{amendt2007best, matuszewski2021post}. These measurements should last from 3 to 10 consecutive days \cite{amendt2007best, johnson2012experimental, hofer2020estimating, lutz2020stay}. The meteorological station should be located within a 15 km radius of the death scene \cite{johnson2012experimental, hofer2020estimating}, and the average differences between the data from the station and the measurements should not exceed 5$^{\circ}$C \cite{johnson2012experimental}. After completing the measurements, the regression analysis should be conducted on the data regressing local measurements against station measurements; typically employing linear regression \cite{archer2004effect, johnson2012experimental, hofer2020estimating, matuszewski2019post}. Some authors tested more complex statistical models, which we review below.

Jeong et al. \cite{jeong2020extended} tested eight different statistical models (linear, quadratic, robust M, least median squares [LMS], leas trimmed squares [LTS], Loess, generalized additive models [GAM], and support vector machines [SVM]) in both indoor and outdoor conditions. Their main purpose was to demonstrate that reconstructing the temperature at a death scene requires using more variables than just local temperature measurements. They propose incorporating wind volume, wind speed, humidity, rainfall, season, and time of measurement. As a result, the mean absolute error decreased by about one degree compared to simple linear regression. However, using so many explanatory variables is impractical. The authors also showed that using only temperature measurements, the error reduction compared to linear regression does not exceed 0.3 degrees on average regardless of the model used.

Moreau et al. \cite{moreau2021honey} applied a GAM model to reconstruct the temperature in a container at a death scene using data on the temperature inside the container versus the temperature outside the container. However, one of the steps in their protocol involved reconstructing the temperature at a death scene using data from a meteorological station. There was no improvement in using the GAM model compared to simple linear regression.

Lutz and Amendt \cite{lutz2020stay} also compared performance of GAM and linear regression reconstructing temperature on the death scene. They tested the effect of the length of measurement period on absolute error of reconstruction. The error of GAM model appeared to be more stable than linear one and it was smaller by approximately 1-2 degrees for measurements lasting 1-5 days and by about 0.1-1 degree for measurements lasting 7-10 days compared to linear model.  

Despite clear recommendation to correct weather station temperature data, review of forensic entomology cases \cite{lutz2020stay} revealed that it was rarely applied in forensic practice, occurring only in approximately 13\% of indoor cases and in approximately 6\% of outdoor cases. Using uncorrected data from meteorological station for outdoor conditions is considered reasonable by some authors \cite{dourel2010using, dabbs2010caution, dabbs2015should}. However, temperatures in quasi-indoor conditions (e.g. uninhabited buildings, garages, or various types of containers) are substantially different to temperatures recorded in the metrological station \cite{michalski2018thermal, lutz2020stay, moreau2021honey}, so uncorrected data are in most cases insufficient, especially over short periods of measurements \cite{lutz2020stay}.

Thus, possible reason for the rare application of temperature correction is the fact that temperature measurements at a death scene are usually connected with practical difficulties. Data loggers can be destroyed or damaged by animals, flooded, or even stolen. It is necessary to periodically check the loggers throughout the measurement period to minimize such risks. Furthermore, if the body is found on a private property, the consent from the owner is necessary to place the logger for measurements. All these limitations likely result in law enforcement personnel being reluctant to the temperature recordings on a death scene. In our opinion the remedy for this appears to be the substantial shortening of measurement period. Unfortunately, according to the majority of authors, current methods are not suitable for correcting temperatures using measurement periods shorter than 5 days \cite{amendt2007best, hofer2020estimating, lutz2020stay} or shorter than 10 days \cite{johnson2012experimental}. Hence, the major aim of this study was to develop an efficient method for reconstructing temperature at the quasi-indoor death scene that requires minimal effort in terms of the measurement duration. 


\section{Materials and Methods}

\subsection{Materials}
The aim of the field temperature recording study was to collect data which were then used to test algorithms proposed in this paper.

Using HOBO Pro v2 2x6’ Ext Temp temperature data loggers model U23-003 (HOBO, USA), we conducted measurements at eight locations in northern Poland, in the village of Skórka (53$^{\circ}$ 13’, 16$^{\circ}$ 52’).

Measurements took place on the roof of a heated building, in the attic of a heated building, in an unheated wooden shack, in an unheated garage being a part of a heated building, in an uninhabited building, and about 30 centimeters below the ground surface. In all locations, two data loggers (with two probes each) were placed in a shaded spots and if possible 30 cm above the ground surface. Throughout the measurement period, no one had access to the data loggers.

In all locations, the measurement period lasted from August 2nd to August 17th, 2021. In each place, there were four probes recording temperature every minute in the same time. For the analyses, we used the average hourly temperature (calculated from all four probes) to align the data with that from the meteorological station.

Data from the station were obtained from the public database of  the Polish Institute of Meteorology and Water Management (\href{https://danepubliczne.imgw.pl/data/dane_pomiarowo_obserwacyjne}{\textcolor{blue}{\underline{link}}}). We used data from the station in Piła that is located approximately 10 km from the study site. The dataset included average hourly temperatures.

\begin{figure} 
	\includegraphics[width=\linewidth]{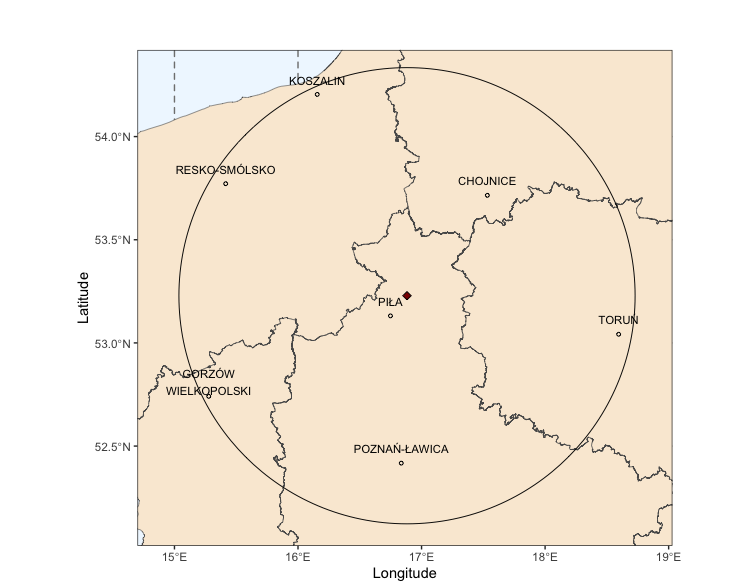}
	\caption{The location of meteorological stations. The red square indicates the study site. The black circle has 125 km (about 78 miles) radius and place of measurements as the center.}
	\label{fig:fig1}
\end{figure}

To check the assumptions of our model, we also used data from the other meteorological stations located within a 125 km radius of place of measurements (Fig. \ref{fig:fig1}).

\subsection{Methods}
\textbf{Train-test split:} To obtain unbiased error estimators for the models, we divided the data into a training set and a test set \cite{xu2018splitting}. To simulate casework conditions, we used the three "oldest" days (August 2nd, 3rd, and 4th) as the test set, while the rest of the data were used as the training set. \\

\noindent \textbf{General idea} \\

\noindent \textbf{Mid-term model:} By "mid-term model", we refer to a statistical model for which the training dataset included measurements, which lasted more than one day. We grouped the data by days. In each group, we had 24 points (one mean temperature for every hour), then we transform these 24 points into a function, both for data from measurements and from meteorological station. In this way, we obtained functional data (i.e. consisting of functions). Afterwards, we used the concurrent regression model, which is the generalization of linear regression to functional data \cite{ramsaysilverman}, where a set of functions (instead of points) for the meteorological station data is the explanatory variable, and a set of functions for the death scene data is the response variable. In addition, coefficients of this model are not numbers but functions – coefficient function. Technical details are described later in this section (point: Technical details for the mid-term model). From now on, we refer to the mid-term model as MTM (Mid-Term Model).

\textbf{Short-term model:} By "short-term model", we refer to an analytical model for which the training dataset included measurements, which lasted less than one day. So, we had less than 24 points. First, using the Bayes’ theorem we found function which was the most probable extrapolation of these points over the entire day. Next, we noticed that percentage differences between subsequent days in characteristics of temperature (e.g. amplitude) are similar in all meteorological stations (Fig. \ref{fig:fig1}) and in all places of measurements. In effect, using a single function from a given place and percentage differences from the nearest meteorological station we were able to derive functions for the other days in the given place by multiplication. Technical details are described later in this section (point: Technical details for the short-term model). From now on, we refer to the short-term model as STM (Short-Term Model).

\textbf{Reference model:} For both tested models we adopted simple linear regression model (hereafter LM) as a reference. The mean absolute error of linear regression model was taken as a reference point, and the goal was to find a model with the lower error. We also tested the GAM model, but similarly to Moreau et al. \cite{moreau2021honey}, it reduced to LM in our case. \\

\noindent \textbf{Technical details for the mid-term model} \\

\noindent The analytical form of the MTM is as follows:
\begin{equation}
    Y(t)=\beta_0(t)X(t)+\beta_1(t)+\epsilon(t),
	\label{eq:regression}
\end{equation}
where $X$ is the explanatory function, $Y$ is the response function, $\beta_0$ and $\beta_1$ are the functions of coefficients, and $\epsilon$ is a function of random error. Respectively, $Y(t)$, $X(t)$, $\beta_0(t)$, $\beta_1(t)$, $\epsilon(t)$ are the values of these functions at the time point $t$. Operations in the formula \ref{eq:regression} are understood as pointwise operations.
In our case, $X:T \rightarrow R$ and $Y:T \rightarrow R$, where $T=[0,24)$ is a set of time over one day, $X$ is a stochastic process representing the temperatures from meteorological station, $Y$ is a stochastic process representing the temperatures from death scene.

\begin{figure} 
	\includegraphics[width=\linewidth]{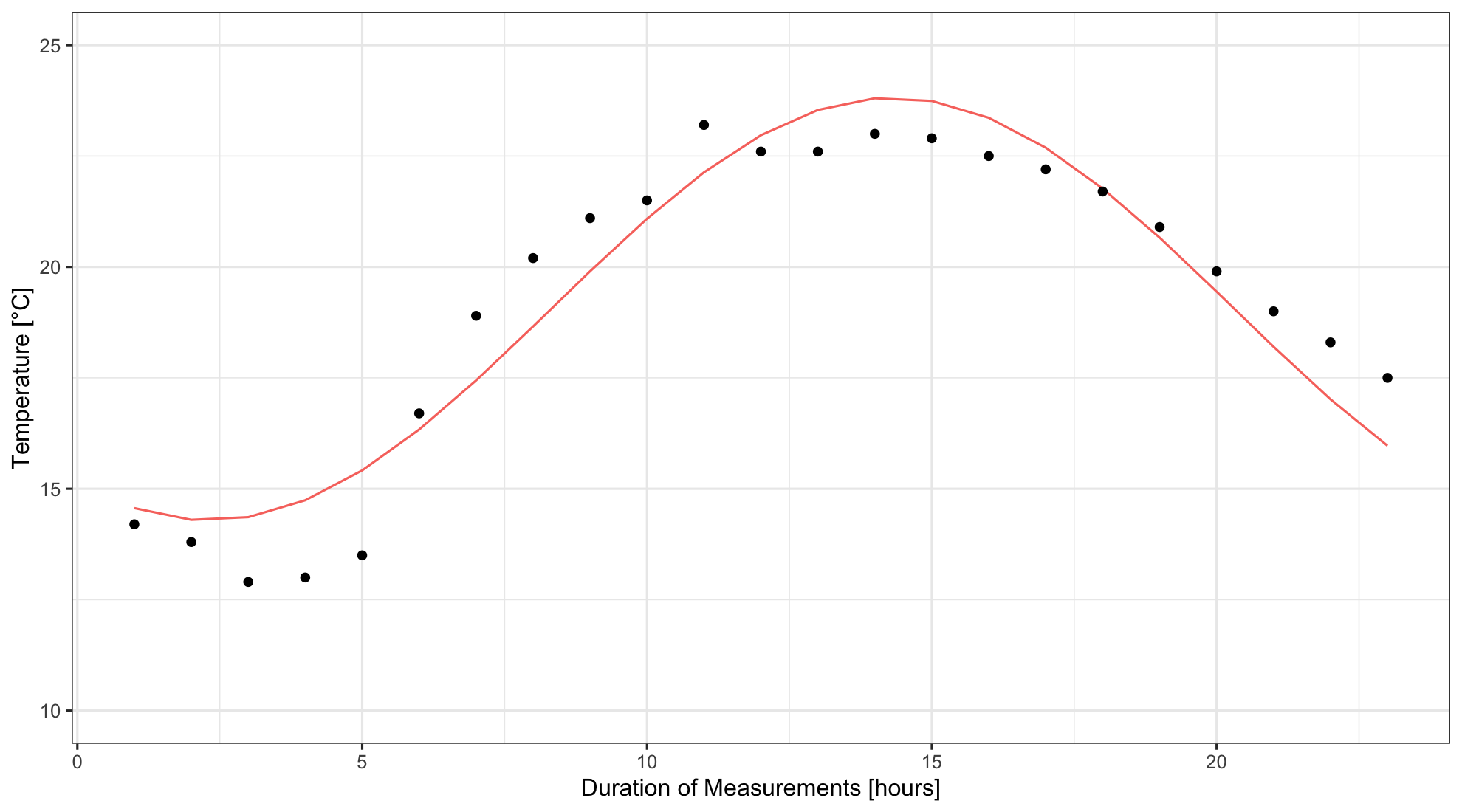}
	\caption{Illustration of data transformation. Black dots represent measurement data (e.g., from a meteorological station), and the red line is a function obtained by smoothing the point data. In this figure, one functional observation was obtained from 24 point observations.}
	\label{fig:fig2}
\end{figure}

\textbf{Data preparation:} The explanatory functions were obtained by smoothing the data from the meteorological station into a truncated Fourier series (Fig. \ref{fig:fig2}):
\begin{equation}
    a_0 + a_1sin(x) + a_2cos(x).
    \label{eq:model}
\end{equation}
We used an algorithm based on QR-decomposition, which is implemented in the `smooth.basis` function from the fda package \cite{ramsay2014fda} in the R language \cite{rcore} with `method` argument set to ‘qr’, i.e. `smooth.basis(…, method = ‘qr’)`. QR-decomposition is a decomposition of a matrix $A$ into a product $A = QR$ of an orthonormal matrix $Q$ and an upper triangular matrix $R$ \cite{francis1961qr, francis1962qr}.

We chose a three-element Fourier basis for two reasons. First, preliminary tests showed that it performs the best. Second, based on a known trigonometric identity, such a series simplifies to a sine function, which aligns with the suggestion by Higley and Haskell \cite{higley2001insect} to represent daily temperature data as a sine wave.

The response functions were obtained in a similar manner by applying the `smooth.basis` function to the data from measurements at the death scene.

\textbf{Regression analysis:} The coefficient functions were obtained using the fda package \cite{ramsay2014fda} using `fRegress` function. These functions also came from a space with a three-element Fourier basis. \\

\noindent \textbf{Technical details for the short-term model} \\

\noindent To reconstruct the function $M(x) = a_0 + a_1sin(x) + a_2cos(x)$ for measurements lasting less than one day, we utilized Bayes' theorem:
\begin{equation}
    p(\sigma|D,M) \propto p(\sigma)p(D,M|\sigma),
    \label{eq:theorem}
\end{equation}
where $p(\sigma|D,M)$ is the posterior probability density function of the model’s residuals' standard deviation, $p(\sigma)$ is the prior probability density function of the model’s residuals, and $p(D,M|\sigma)$ is the likelihood, with $D$ denoting the measurement data.

We assumed that $p(\sigma)=Norm(\sigma,1,0.75)$ and $p(D,M|\sigma) = \Pi_xNorm(resid(x), 0, \sigma)$, where $Norm$ denotes the density of the normal distribution with the mean and standard deviation as the second and third arguments, respectively; $resid(x)=(y-M(x))$ is the model residual at a point $x$ ($y$ is observation in point $x$); and $\sigma$ is the standard deviation of the residuals for a given model.

The parameters of the prior distribution and likelihood were determined experimentally for our data loggers, prior to this article. We measured temperature for a few days in various locations (the scheme was the same as in the main experiment), then we transformed point data to functional data using `smooth.basis` function and we measured characteristic of residuals. In every location residuals were approximately normally distributed with mean $0$ and different standard deviation. Afterwards, we examined distribution of these standard deviations and it occurred that they are also normally distributed with mean close to one and standard deviation close to three-quarters. So we took this information as assumptions to our Bayesian reasoning.

Subsequently, we selected the model $M$ for which $p(\sigma|D,M)$ is the largest. In order to achieve this, we searched a grid of parameters that met the conditions:
\begin{equation}
    t_m-5<a_0<t_m+5, 
    \label{eq:cond1}
\end{equation}
with step 0.25,
\begin{equation}
    \sqrt{a_1^2+a_2^2}<20, 
    \label{eq:cond2}
\end{equation}
with step 0.01, where $t_m$ is a mean temperature from meteorological station. We assumed that the daily temperature amplitude in Central Europe during the summer cannot exceed $20$ degrees Celsius, and the average temperature at any of our locations do not deviate by more than five degrees from the average temperature at the meteorological station.

In this way, we obtained an approximation of the temperature function for the entire day in which the measurements were made.
To obtain functions for other days, we transformed data from meteorological station into functional data using `smooth.basis` function like in the case of mid-term model. Let $b_i^{(j)}$ be the value of the coefficient $b_i$ (corresponding to the coefficient $a_i$) calculated for the data from the meteorological station on the $j$th day. Let $h_i(j)=\frac{b_i^{(j)}}{b_i^{(0)}}$, then we assumed that:
\begin{equation}
    a_i^{(j)}= h_i(j) a_i^{(0)}
    \label{eq:ass1}
\end{equation}
where $a_i^{(j)}$ is the value of the corresponding coefficient $a_i$ on the $j$th day. We also assumed that we can calculate function $h_i$ using data from any meteorological station from Figure \ref{fig:fig1} or any place of measurements and its values are very similar. We refer to this assumption as: the relative differences in coefficient values remain approximately the same from day to day.
The statistical test for these assumptions can be found in the Results section.\\

\noindent \textbf{Performance measurements} \\

\noindent To measure the errors of the models, we used the Mean Absolute Error (MAE):
\begin{equation}
    MAE = \frac{1}{n}\sum_{k=1}^n|y_i - f(x_i)|
    \label{eq:mae}
\end{equation}
where $y_i$ is the measurement at $x_i$, and $f(x_i)$ is the value of the model at $x_i$.
We did not use the coefficient of determination $R^2$ because it is applicable only to linear models \cite{spiess2010evaluation}, and both MTM and STM are nonlinear models.\\

\noindent \textbf{Statistical tests} \\

\noindent To test the assumption about relative differences in coefficients of functional data $a_j$ between subsequent days (see the Short-term model subsection) we used Mean Absolute Error (MAE) and the paired t-test (details below).

We calculated relative differences in coefficients of functional data ($a_i^{(j)}$) between subsequent days at each location where we took measurements as well as at the meteorological station ($b_i^{(j)}$). Then, we calculated the MAE between each location and the meteorological station $\frac{1}{n}\sum_{k=1}^n|y_i - f(x_i)|$, resulting in 18 values (3 coefficients across six locations) to assess whether the differences were high.
Similarly, using paired t-test we checked if differences between $(a_i^{(1)},a_i^{(2)},…,a_i^{(n)})$ and $(b_i^{(1)},b_i^{(2)},…,b_i^{(n)})$ were significant. This was done for each coefficient and each location, leading to a total of 18 tests.
We also applied a one-way ANOVA test to check for significant differences between corresponding data from all meteorological stations shown in Figure \ref{fig:fig1}. We operated on smoothed data using the `smooth.basis` function. In other words, we checked if $(b_i^{(1)},b_i^{(2)},…,b_i^{(n)})$ differs between stations.\\

\noindent \textbf{Software} \\

\noindent All analyses were performed in R (version 4.3.2) using the RStudio IDE, with the packages: fda \cite{ramsay2014fda} and the tidyverse ecosystem \cite{wickham2019welcome}. Calculations were conducted on an Apple MacBook Pro with an M1 processor in the aarch64-apple-darwin20 architecture.


\section{Results}
\subsection{Test for assumption about relative differences}
The relative differences in coefficients at the meteorological station (Piła) are presented in Table  \ref{tab:tab1}.

\begin{table*} 
    \caption{Relative differences in coefficients of data for different days at the meteorological station expressed as ratios of coefficients for the given day and the day of measurements}
    \centering 
    \begin{tabular}{cccccccccccccccc}
        \toprule
        \multicolumn{15}{c}{The day of measurement} \\
        \cmidrule(r){2-16}
         & 0 & 1 & 2 & 3 & 4 & 5 & 6 & 7 & 8 & 9 & 10 & 11 & 12 & 13 & 14 \\
        \midrule 
        $b_1$ & 1 & 1.01 & 1 & 1.09 & 0.94 & 1.06 & 1.02 & 0.96 & 1 & 0.90 & 1.10 & 1.17 & 0.95 & 0.97 & 1.02 \\
        $b_2$ & 1 & 1.27 & 1.85 & 1.23 & 0.39 & 4.54 & 0.58 & 0.89 & 0.63 & 1.41 & 1.95 & 1.12 & 0.36 & 1.52 & 1.04 \\
        $b_3$ & 1 & 1.04 & 0.59 & 1.35 & 1.13 & 1.48 & 0.63 & 1.28 & 1.29 & 0.71 & 1.49 & 1.09 & 0.47 & 1.25 & 1.73 \\
        \bottomrule
    \end{tabular}
    \caption*{In the column zero, there is the ratio between the coefficient $b_i$ recorded on the day of measurement and itself; in the column one, the ratio between the coefficient $b_i$ recorded the day after the day of measurement and the coefficient $b_i$ recorded on the day of measurement; in the column two, the ratio between the coefficient $b_i$ recorded on the second day after the day of measurement and the coefficient $b_i$ recorded on the day of measurement, and so on. According to our assumption, these ratios are equal to the ratios of respective $a_i$ coefficients. Coefficients $b_i$ are recorded at the meteorological station, while coefficients $a_i$ are coefficients from the death scene.}
    \label{tab:tab1}
\end{table*}

The relative differences in coefficients between each measurement location and the meteorological station were not statistically significant. Paired t-tests were conducted separately for each of the three coefficients at each of the six locations, resulting in a total of 18 tests, with all p-values greater than 0.05. The Mean Absolute Error (MAE) also indicated that the measurements across locations were very similar. Assumptions for the paired t-tests were checked, and the results confirmed that they were met.

Similarly, no significant differences were observed in the coefficients recorded across the meteorological stations (one-way ANOVA, one test per coefficient, a total of three tests, with all p-values greater than 0.07). The assumptions for ANOVA were also verified, and the results indicated that they were satisfied.

Thus, the assumption that relative differences in coefficients do not vary significantly from day to day is supported.

\subsection{Mid-term reconstruction}
\begin{figure*} 
	\includegraphics[width=\linewidth]{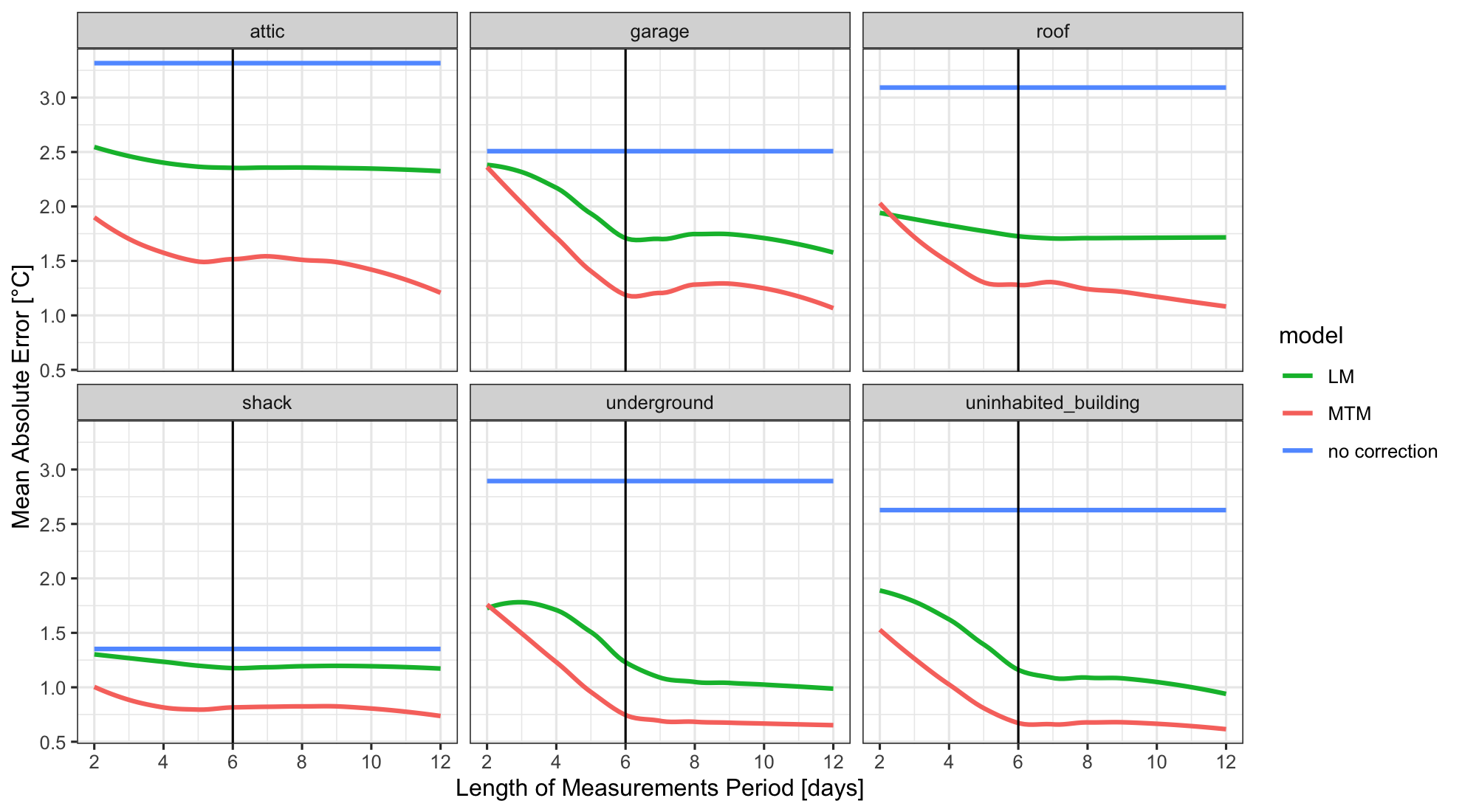}
	\caption{Correction errors (in mean absolute errors, MAE) when MTM (red) and linear regression (green) were used to model the temperature in tested places (attic, garage etc.) under various measurement periods [in hours]. Blue line represents the error of data from meteorological station with no correction. Black line represents 6-days measurements after which MAEs stabilized. MAEs were calculated on the test set.}
	\label{fig:fig3}
\end{figure*}

\begin{figure*} 
	\includegraphics[width=\linewidth]{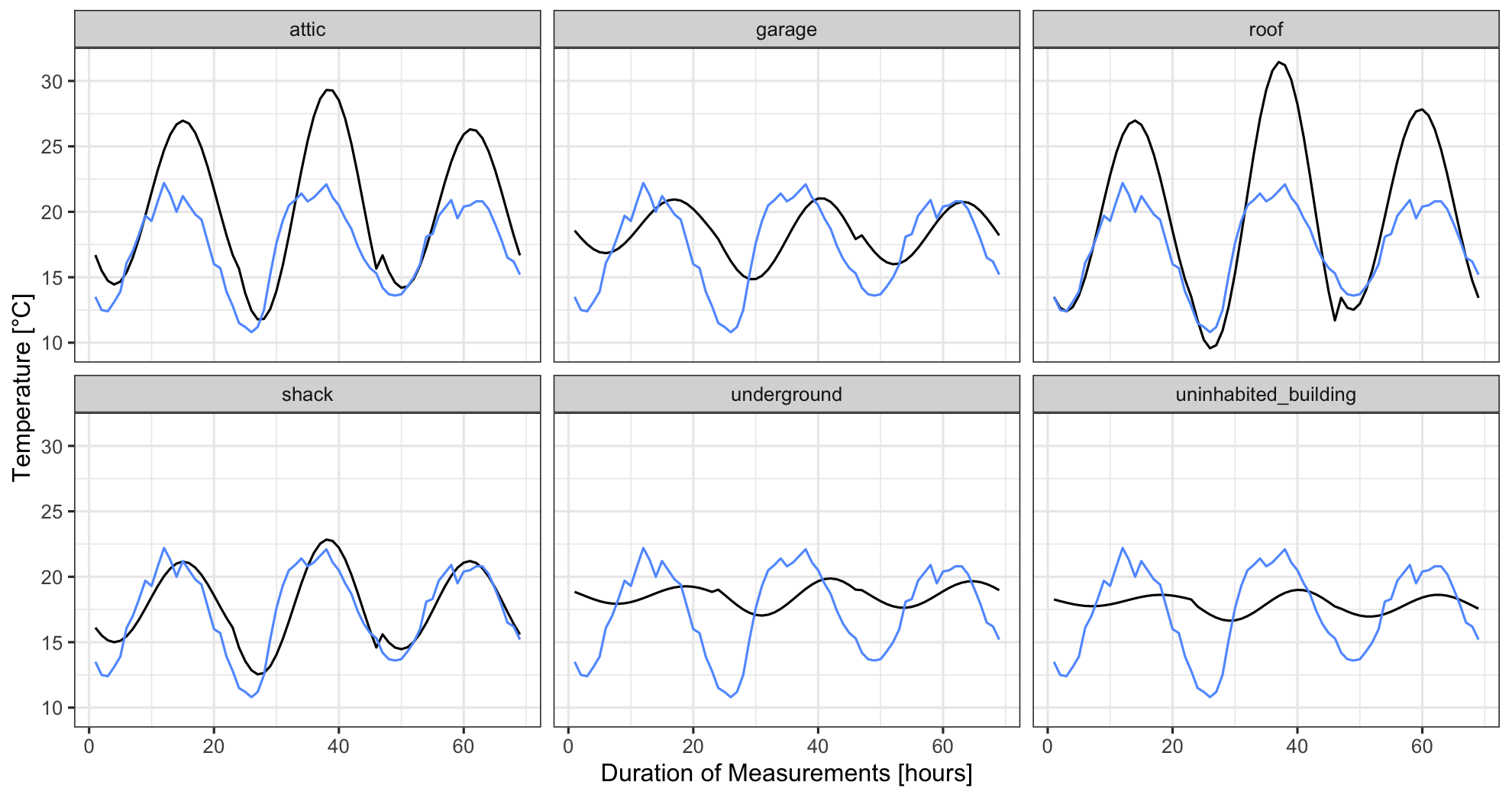}
	\caption{Comparison of uncorrected data from meteorological station (blue) and from death scene (black) using the test set.}
	\label{fig:fig4}
\end{figure*}

\begin{figure*} 
	\includegraphics[width=\linewidth]{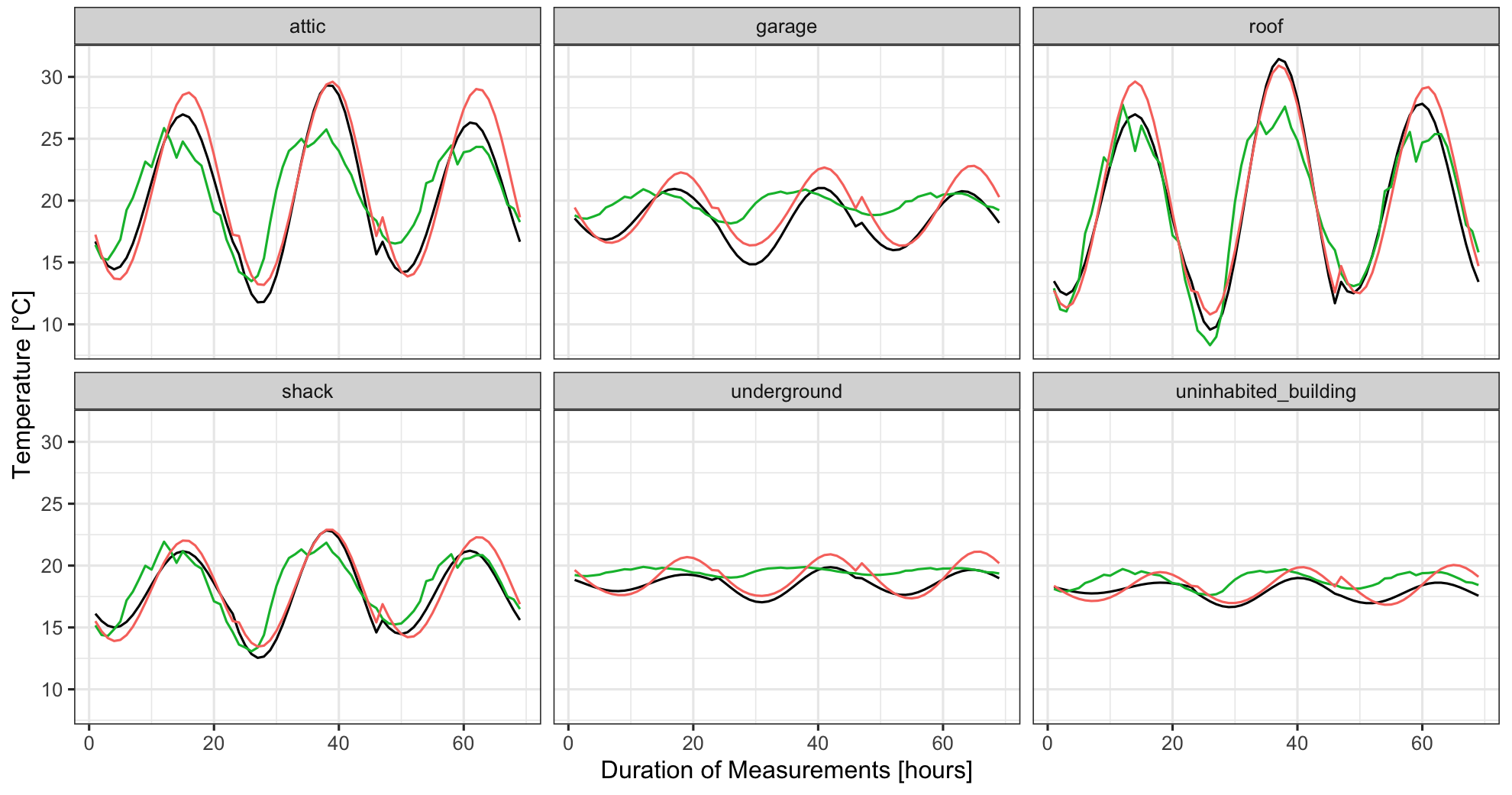}
	\caption{Comparison of data from weather station corrected by linear regression (green), corrected by MTM (red) and from death scene (black) on the test set. Measurements used for the correction lasted 12 days.}
	\label{fig:fig5}
\end{figure*}

When measurements lasted more than one day, the MTM resulted in nearly twice smaller error than in the case of the linear regression model. After six days, the error of the MTM stabilized and the temperature reconstruction was nearly perfect (Fig. \ref{fig:fig3}). Uncorrected data from the meteorological station did not match the measurements (Fig. 3). In all locations, there was a nonzero phase shift, indicating that minima and maxima at the meteorological station were reached at different times of a day than at the death scene (Fig. \ref{fig:fig4}). Correction of temperatures using linear regression improved the correspondence between the death scene and station temperatures (Fig. \ref{fig:fig5}), but phase shifting remained a problem, especially in locations where temperatures were less variable (e.g. uninhabited building or garage). Although the MTM did not provide a perfect correction, it was much better than the linear model (Fig. \ref{fig:fig5}).

\subsection{Short-term reconstruction}
\begin{figure*} 
	\includegraphics[width=\linewidth]{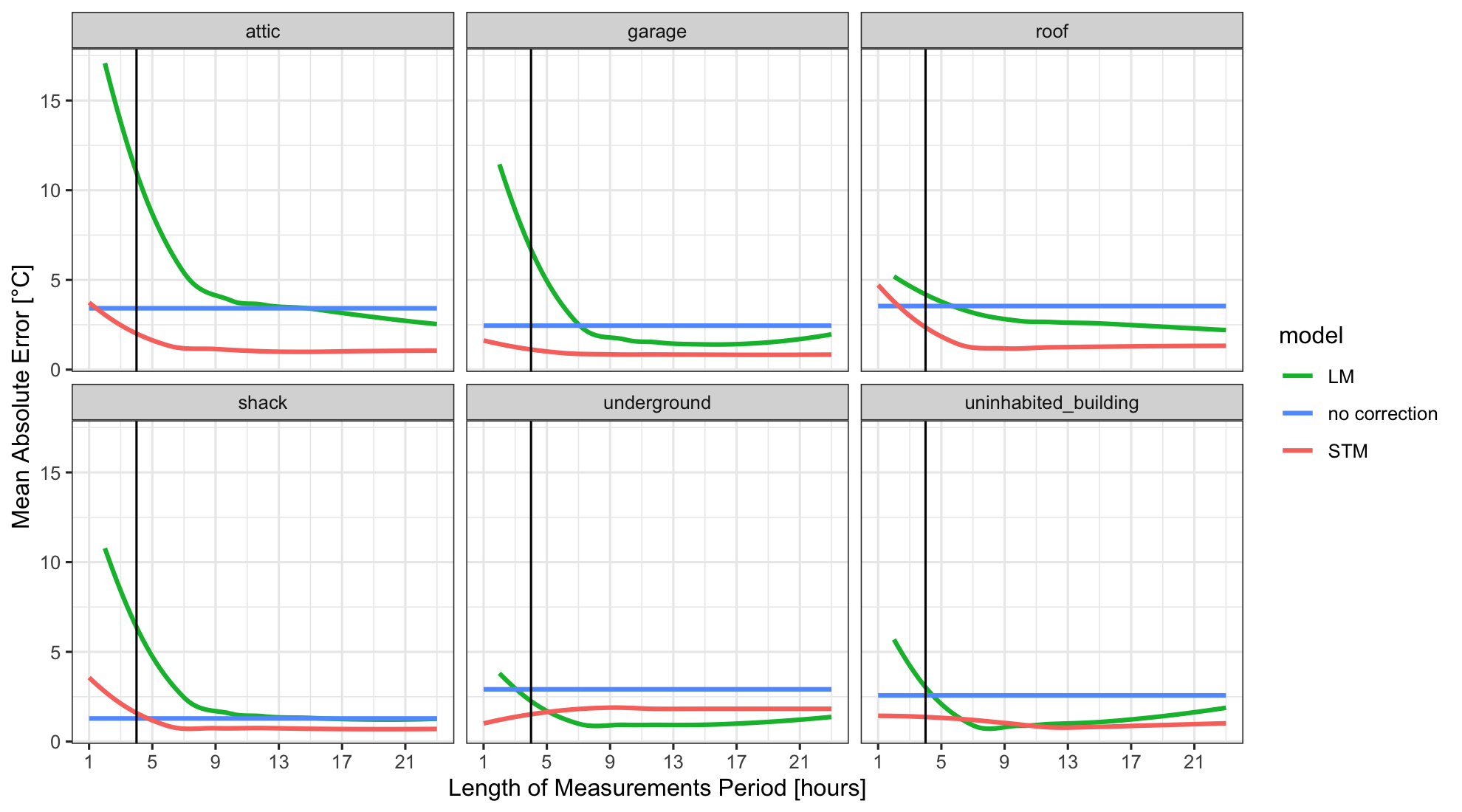}
	\caption{Correction errors (in mean absolute errors, MAE) when STM (red) and linear regression (green) were used to model the temperature in tested places (attic, garage etc.) under various measurement periods [in hours]. Blue line represents the error of data from meteorological station with no correction. Black line represents 4-hours measurements after which MAEs stabilized. MAEs were calculated on the test set.}
	\label{fig:fig6}
\end{figure*}

\begin{figure*} 
	\includegraphics[width=\linewidth]{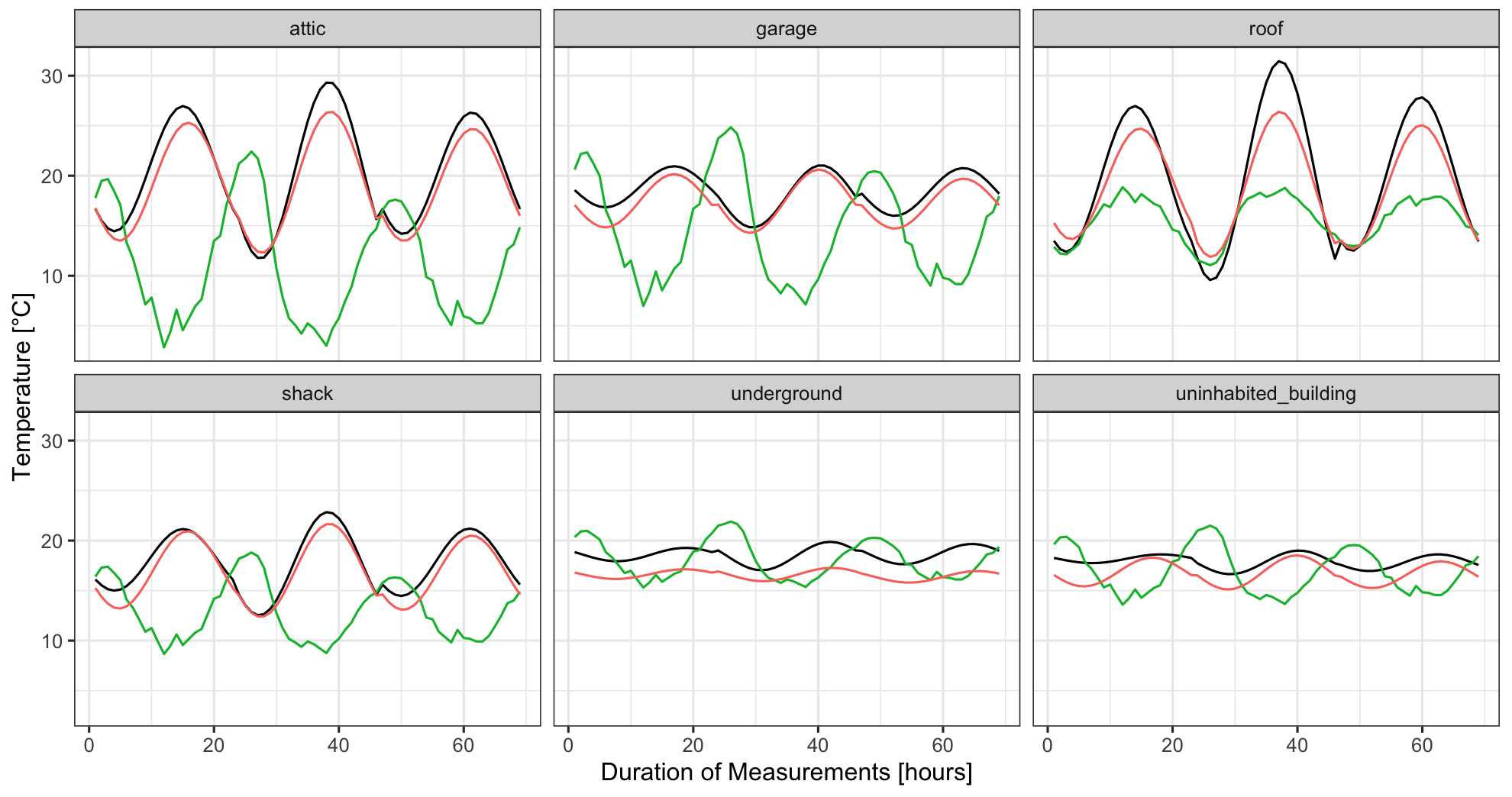}
	\caption{Comparison of data from weather station corrected by linear regression (green), corrected by STM (red) and from death scene (black) on the test set. Measurements used for the correction lasted 4 hours.}
	\label{fig:fig7}
\end{figure*}

When measurements lasted several hours, in five study places the linear regression yielded worse results than using the uncorrected data from the meteorological station (Fig. \ref{fig:fig6}). In the same five places (i.e., attic, garage, roof, shack, and uninhabited building), data reconstruction with the STM returned better results. After about 4-5 hours, the error rate became similar to the error rate achieved by the MTM after six days (Fig. \ref{fig:fig3} and 6). Although the numerical error for underground measurements was slightly lower using linear regression, temperatures corrected with STM accurately reproduced minima and maxima and only the average temperature was underestimated (Fig. \ref{fig:fig7}).

\section{Discussion}
From the perspective of thermal characteristics, death scenes can generally be divided into three groups: indoor, outdoor, and other scenes \cite{michalski2018thermal}. In indoor scenes, the temperature is usually fully controlled by humans or appropriate machine and remains most of the time relatively constant. It typically ranges from about 20 to 30 degrees Celsius \cite{peeters2009thermal, van2010thermal, yang2014thermal} depending on the resident’s preferences. Consequently, the reconstruction of temperatures in indoor death scenes can be based simply on measurements taken during the inspection of the scene, the knowledge of the residents’ preferences and the heating devices present inside (their operation and settings).

The temperature in outdoor death scenes should be close to the temperature recorded at meteorological station. First, this aligns with the primary purpose of meteorological stations, which is to provide information about outdoor temperatures. Second, this assumption is supported by some studies on the reconstruction of temperatures in outdoor death scenes \cite{dabbs2010caution, dabbs2015should, charabidze2019temperature}, in case of which authors conclude that the reconstruction is unnecessary. Moreover, a closer examination of the studies, in which the reconstruction was recommended, even for outdoor death scenes, reveals that the benefit is usually rather minor \cite{archer2004effect, johnson2012experimental, hofer2020estimating}.

Death scenes that cannot be classified as indoor or outdoor (referred in this work as quasi-indoor, e.g. attics, garages, shacks or uninhabited buildings), frequently exhibit a specific microclimate resulting from differences in heat flow due to the partial insulation or other thermal barriers and lack of heating devices \cite{lutz2020stay}. As a result, the temperature on such scenes significantly differs from that recorded at meteorological station, yet it is not constant \cite{michalski2018thermal}.

Death scenes such as attics are separated from outdoor conditions by the barrier of a roof (frequently uninsulated) additionally, they are influenced by heat from the living area. However, usually they are unheated, so the temperature in such places is less controlled than in ordinary rooms.  Places such as garages, shacks or uninhabited buildings are isolated from outdoor conditions but usually do not have own heating devices. Therefore, the temperature inside such places reveals typical daily fluctuations, however with limited daily amplitude as compared to outdoor conditions. At the same time, the extent of insulation in each of these places can differ (e.g. thin walls in a shack or thick walls in an uninhabited building), in effect substantial thermal differences can be expected across these scenes. All these peculiarities of quasi-indoor death scenes indicate that correction of weather station temperatures may be crucial in such places \cite{matuszewski2019post}.

Given these diverse thermal conditions, it is crucial to conduct measurements in such locations and compare them with meteorological data to test the method proposed in this paper and ensure that they reconstruct temperature accurately. In this study, measurements were conducted within a distance of less than 15 km from a meteorological station, meeting the literature’s requirements \cite{johnson2012experimental, hofer2020estimating}. However, in a few cases the average differences between the data from the station and the place of measurements exceeded 5°C, what contradicted another requirement from the literature \cite{johnson2012experimental}. However, these differences had no effect on the quality of our reconstructions. This is an advantage of current models; they are simply resilient to large temperature deviations.

For the mid-term reconstruction, errors stabilized after 6 days, which is in line with the recommendation that measurements at the death scene should last from 3 to 10 days \cite{amendt2007best, johnson2012experimental, hofer2020estimating, lutz2020stay}. The error of the MTM is much lower than that of the linear regression model. Comparing our results with those previously published, errors of the MTM were lower than these of the GAM \cite{lutz2020stay, moreau2021honey} or SVM \cite{jeong2020extended} models, when only one explanatory variable was used and they were comparable to the situations where more variables were used \cite{jeong2020extended, moreau2021honey}. It seems that better results are hardly obtainable. However, the error of the linear regression model in this study was higher than in the previous studies \cite{jeong2020extended, lutz2020stay, moreau2021honey}. Since previous authors probably did not divide the data into training and test sets, this could be the reason for their lower errors. Evaluating errors with the same data that were used to train the model usually results in an underestimation of the error values \cite{xu2018splitting}. However, these differences might also stem from other sources (e.g. different places of the studies or different equipment used to record temperatures).

The MTM yielded lower errors compared to linear regression, GAM or SVM models. Moreover, it correctly reconstructs temperature characteristics such as the positioning of minima and maxima, amplitude, and average temperature. These properties are poorly reconstructed by the linear regression model, while GAM and SVM require more explanatory variables, which is impractical.

The STM allowed for the substantial reduction of measurement period. Thanks to this model, reconstruction of temperatures based on measurements lasting a few hours resulted in an error level close to that achieved by the best models after several days of measurements and without the need to use additional explanatory variables. This is a significant novelty of the current work. As the reluctance to use temperature reconstruction techniques by law enforcement staff shown by Lutz and Amendt \cite{lutz2020stay} probably stems from the necessity of conducting long-term measurements, current proposal can have also important practical implications. The technique proposed in this paper allows reducing measurements period to about five or even four hours. In consequence, the practical costs of recording temperatures on a death scene can be reduced to negligible values, as it would suffice to lay out a data logger at the beginning of the death scene examination and record data only for the duration of the examination \cite{amendt2007best}. The device would then be relatively safe, not exposed to risks such as flooding, theft, or damage by animals which are quite high when measurements last for several days. This could remove one of the main reasons, due to which the temperature reconstruction is rarely used in practice \cite{lutz2020stay}.

However, the STM exhibits several significant drawbacks. First, there is a need of knowing several characteristics of the data loggers, such as their typical residual distribution. This means that the device used for measurements must be tested beforehand. Alternatively, the expert should be allowed to test the device later. Second, a one-time test of a data logger assumes no time-dependent variation in residual distribution. Since for most devices this is probably not true, such tests should be performed on a regular basis (similar to their calibration tests, \cite{amendt2007best, girondot2018understanding}.

Third, the algorithms of both presented models are difficult to implement, this difficulty is particularly high in the case of the algorithm of STM. While its calculation procedures are not complex, there is currently no software available for their execution using computers. Consequently, non-technical users may require assistance from mathematicians or data scientists. The situation with the MTM is slightly more favorable, with ready-to-use packages accessible for R or Python languages \cite{ramsaysilverman, ramsay2014fda, ramos2022scikit}, although it still demands some programming proficiency. However, developing specific software for both models is not challenging but does require additional funds. Our R-scripts are available from the corresponding author on a reasonable request.

We recommend using the MTM when experts can make measurements for at least six days. Otherwise, we suggest using the STM and perform measurements for at least four hours (preferably five to six hours), ideally during the death scene examination, to minimize the costs.

Faced with choosing between MTM and STM, one should opt for MTM. Although both models reproduce the amplitude and positioning of temperature minima and maxima with similar accuracy, the MTM slightly better reconstructs the average daily temperature (e.g. as in the uninhabited building or underground in this study). The STM may also less accurately (than MTM) reproduce the amplitude (e.g. attic and roof in this study).

\subsection{Declaration of generative AI and AI-assisted technologies in the writing process}
During the preparation of this work, the authors used ChatGPT to assist with grammar improvements. Following the use of this service, the authors reviewed and edited the content as necessary and take full responsibility for the final version of the manuscript.

\subsection{Authors' Contributions}
All authors contributed to the conceptualization, review, and editing of the manuscript, and approved the final version of the text. Jędrzej Wydra was primarily responsible for programming, investigation, data collection and organization, formal and data analysis, validation, visualization, and drafting the original manuscript. Łukasz Smaga and Szymon Matuszewski provided supervision.

\subsection{Compliance with ethical standards}
This paper does not involve any studies with human participants or animals conducted by any of the authors.

\subsection{Funding}
The authors received no specific funding for this study.




\end{document}